\documentclass[12pt,preprint,psfig]{aastex}
\usepackage{psfig}
\slugcomment{Draft}

\shorttitle{Ejecta Structure of Cas A}
\shortauthors{Laming \& Hwang}

\begin{document}
\title{On the Determination of Ejecta Structure and Explosion Asymmetry
from the X-ray Knots of Cassiopeia A}


\author{J. Martin Laming\altaffilmark{1} \& Una Hwang\altaffilmark{2}}


\altaffiltext{1}{Code 7674L, Naval Research Laboratory, Washington DC 20375\\
\email{jlaming@ssd5.nrl.navy.mil}}
\altaffiltext{2}{Goddard Space Flight Center and University of Maryland\\
\email{hwang@orfeo.gsfc.nasa.gov}}

\begin{abstract}
We present a detailed analysis of Chandra X-ray spectra from
individual ejecta knots in the supernova remnant Cassiopeia A. The
spectra are fitted to give the electron temperature $T_e$ and
(single) ionization age $n_et$. These quantities are compared with the
predictions of self similar hydrodynamic models incorporating time
dependent ionization and radiation losses, and Coulomb electron-ion
equilibration behind the reverse shock, for a variety of different
ejecta density profiles described by a uniform density core and a
power law envelope. We find that the ejecta close to the ``jet''
region in the NE, but not actually in the jet itself, have a
systematically shallower outer envelope than ejecta elsewhere in the
remnant, and we interpret this as being due to more energy of the
initial explosion being directed in this polar direction as opposed to
equatorially. The degree of asymmetry we infer is at the low end of
that generally modelled in asymmetric core-collapse simulations, and
may possibly be used to rule out highly asymmetric explosion models.
\end{abstract}
\section{Introduction}
It has become reasonably well accepted that the star recorded in the
astronomer Flamsteed's 1725 catalogue as 3 Cassiopeiae was indeed the
supernova that gave rise to the remnant now known as Cassiopeia A
\citep{ashworth80}. Flamsteed's observation in 1680 makes Cas A one of
a handful of historical supernova remnants for which we have a precise
age. In fact Cas A is by now well studied by astronomers at all
wavelengths from the radio \citep[it is the brightest source in the
northern sky,][]{reber44,ryle48} to TeV $\gamma$-rays
\citep{aharonian01}, allowing determinations with a high degree of
confidence of important parameters such as the expansion velocities of
various parts of the remnant \citep{fesen01} and its distance
\citep{reed95}.  This makes it an attractive ``laboratory'' for
studying various physical phenomena, such as the plasma physics
connected with particle acceleration \citep{vink02}, or
nucleosynthesis through observed element abundances \citep[see
e.g.][]{vink01}.

In this paper we analyze the X-ray emission from small localized knots
of material within the Cas A ejecta in order to try and constrain the
ejecta density profile. The sub-arcsecond spatial resolution of the
Chandra X-ray Observatory has made it possible to study individual
structures in the Cas A shell. These structures have angular dimension
typically $\sim 1''$, which is $\sim 5\times 10^{16}$ cm at the 3.4
kpc distance of Cas A. Consequently a reverse shock of velocity 1000
km s$^{-1}$ (a typical value in our models below) will traverse these
structures in around 20 years, a time significantly shorter than the
evolution time of Cas A. Thus they may be fit with a single value of
the ionization age $n_et$, and models for the evolution of the SNR may
be applied to infer the timing of the shock heating in a similar way
to the use of nuclear reaction rates elsewhere in astrophysics, as a
powerful tool for investigating the structure of Cas A. We emphasise
that our goal is not to try and analyze the complete emission from the
ejecta of Cas A, but that by isolating the X-ray knots we are
focussing on those structures in Cas A for which we think we have the
best physical understanding. In this paper we treat only those knots
that appear to be O-rich, in order to constrain the ejecta density
profile and possible asymmetries in the initial explosion. A companion
paper considers in more detail a set of Fe rich knots on the east
limb, determining element abundances for these knots, and from
estimates of the Lagrangian mass coordinate for each knot, makes an
initial attempt to build up a radial profile of the composition of the
supernova ejecta.

\section{The Circumstellar Density Profile}
The forward shock speed in Cas A has been recently measured from the
two extant Chandra ACIS observations with substantial exposures to be
in the range $\sim 4000 - 6000$ km s$^{-1}$ \citep{delaney03}, with an
average value of 4916 km s$^{-1}$.  This is consistent with previous
estimates based on the observed expansion of the bright ejecta ring
\citep{vink98,koralesky98,gotthelf01}. The average radius is 2.46 pc,
leading to an expansion parameter of 0.654. The fact that the ejecta
are still clearly visible in X-rays has suggested to various authors
that Cas A must be in transition from ejecta dominated to Sedov-Taylor
behavior, and the expansion parameter of 0.654 appears most consistent
with a uniform circumstellar medium, for which it should vary between
$1\rightarrow 0.4$ in these two limits. Simple interpretations of the
spectra of forward and reverse shocked material also indicate that the
masses of emitting ejecta and swept up circumstellar gas are
similar \citep{fabian80}.

Based on optical spectra, however, the ejecta of Cas A are believed to
be rich in O \citep{chevalier79}, and if this were the dominant
constituent of the ejecta, the mass inferred from the thermal
bremsstrahlung emission measure would be lower by approximately a
factor of 4 than if it were of an H or He dominated
composition. \citet{vink96} were the first authors to make this point
with regard to X-ray observations, and inferred an ejecta mass of
$\sim 4 M_{\sun}$. \citet{favata97} revised this mass estimate to the
range $2-4 M_{\sun}$ with further uncertainty arising from how the
spectral component responsible for the hard X-ray emission behaves at
lower (i.e. thermal) photon energies.  Recently \citet{willingale02}
also inferred an ejecta mass of 2.2 $M_{\sun}$. These lower ejecta
masses imply that the Cas A blast wave has swept up a relatively {\em
higher} mass of circumstellar material, dominated in composition by H
and He, with N also over abundant relative to solar values. In this
case we expect Cas A to have evolved to the Sedov-Taylor limit, and
the observed expansion factor would imply a stellar wind solution for
the circumstellar density profile, i.e. $\rho\propto r^{-2}$.  The
existence of a pre-supernova stellar wind is also suggested by the
quasi-stationary flocculi \citep{vandenbergh71}, which are presumed to
have formed from stellar wind material.

Another line of argument pointing to a stellar wind density profile is
from the existence of ejecta at optical emitting temperatures. The
reverse shock encounters the densest ejecta at the core-envelope
boundary, and these are the ejecta most likely to undergo thermal
instability during the evolution of a young supernova remnant.  In the
case of a uniform circumstellar medium, for typical SNR parameters,
the reverse shock only penetrates the ejecta core after a time of
order 100 years, by which time the ejecta have expanded to such an
extent that the radiative cooling time is now much longer than the
remnant age. However with a stellar wind ambient medium, the initial
deceleration of the blast wave sends the reverse shock into the ejecta
core on a timescale of order a few years, when the ejecta are still
dense enough to radiate significantly. This point was first made by
\citet{chevalier94}, and is illustrated in Tables 1 and 2 where the
mass of ejecta (assumed pure oxygen) that can cool to $10^4$ K within
320 years of explosion is given for various ejecta envelope power
laws.  Cas A is a little unusual in that it does indeed have optically
emitting ejecta, both in knots and in nebulosity, again strongly
suggesting an $s=2$ or similar ambient density profile. The existence
of radiatively cooled shocked ejecta may also be relevant to the
free-free absorption observed towards the center of Cas A in the radio
\citep{kassim95,anderson95}. Explanations in terms of absorption by
unshocked ejecta pose difficulties in that they require unphysically
large masses of ejecta. Radiatively cooled shocked ejecta at
temperatures of a few thousand K can be much more dense, hence
dramatically reducing the mass of cool plasma required to give the
necessary absorption, while producing negligible photoelectric
absorption of X-rays. The secular decrease of the radio emission from
Cas A \citep{reichart00} has also been interpreted in terms of the
adiabatic expansion of electrons trapped in a shock wave where the
magnetic field is also secularly decreasing, as one would expect in a
remnant circumstellar wind \citep{berezhko03}.

Weaker arguments in favor of an $s=2$ external density profile are
discussed in \citet{vink02}. The hard X-ray emission out to $\sim 100$
keV photon energy has been interpreted as due to non-thermal
bremsstrahlung from a population of electrons accelerated by lower
hybrid waves generated by secondary shocks within the Cas A shell
\citep{laming01a,laming01b}.  Bremsstrahlung is a relatively
inefficient emission process; much more of the electron energy is
dissipated as heat by Coulomb collisions between accelerated and
ambient electrons. Relative to $s=0$, an SNR expanding into an $s=2$
circumstellar medium expands faster at later times, and this extra
adiabatic expansion allows more heat to be deposited in the plasma
without a large increase in temperature. Additionally, the existence
of thermal instability in certain regions of the ejecta allows a much
higher power loss by radiation.  Consequently, in as far as one is
prepared to accept the model of \citet{laming01a,laming01b},
\citep[and this is discussed at some length in][]{vink02}, a stellar
wind external density profile is much more plausible than a uniform
ambient density. In the same manner, non thermal bremsstrahlung
emission models are much easier to construct in more radiative heavy
element dominated ejecta than in an H or He dominated plasma. More
recently \citet{delaney03} encountered difficulties in trying to
interpret their measurements of the forward shock velocity and radius
in terms of models using constant ambient density from
\citet{truelove99}. Setting the ratio of the forward and reverse shock
radii at approximately 3:2 \citep{gotthelf01}, the models overpredict
the observed free expansion rate (determined from outlying optical
knots). The inferred ejecta mass is also rather small at $1.4
M_{\sun}$. Moving to the $s=2$ models described below (see section
3.3), the $n=9$ models give the correct ratio of the forward and
reverse shock ratio, the correct relationship between the forward
shock and free expansion rates, and a more plausible ejecta mass of $2
M_{\sun}$.
Looking more closely at Figure 2 of \citet{gotthelf01}, the
ratio of forward to reverse shock ratio actually varies, being about
1.5 on the west limb, but larger $\sim 1.8-1.9$ on the east limb,
suggesting a lower value for the ejecta envelope power law in this
region. This observation will be interpreted in more quantitative
detail below.

\section{The Ejecta Density Profile}
\subsection{Are the Ejecta Knots really Knots?}
A significant fraction of the ejecta of Cas A are observed in X-rays
to be in knots or clumps. Some of these are located very close to the
forward shock position, inviting the somewhat naive assumption that
they are significantly overdense compared to their surroundings. We
argue below based on a number of lines of reasoning that these knots
are not significantly overdense, but have high ionization ages due to
encountering the reverse shock early in the evolution of Cas A. Their
apparent positions close to the blast wave may possibly be due to
efficient particle acceleration and stronger resulting shock
compression at the forward shock, as modelled by
\citet{blondin01a}. However we find no other indications in Cas A,
either morphological or spectral, of shock compressions greater than 4
that would then be expected, and we do not pursue this issue
further in this paper.

If these knots did indeed undergo reverse shock passage early in the
evolution of Cas~A, then an important question arises as to how they
survived to be observed as knots by Chandra/ACIS. A number of authors
\citep{wang01,klein94,klein03,mckee75,poludnenko01} have modelled
cloud-shock or knot-shock interactions. Upon entering a higher (lower)
density medium, the shock decelerates (accelerates). The cloud or knot
undergoes a different acceleration upon shock passage relative to the
ambient plasma, which can give rise to a Kelvin-Helmholtz instability
at the interface. In addition the shock inside the cloud or knot will
be split into transmitted and reflected shocks each time it encounters
the cloud/knot boundary with the ambient plasma, giving rise to
further Rayleigh-Taylor or Richtmeyer-Meshkov instabilities. The net
result is the destruction of the cloud or knot on a timescale of a few
shock crossing times. For knots $1''$ across in Cas A ($5\times
10^{16}$ cm) the shock crossing time is 15$-$30 years for a 500$-$1000 km
s$^{-1}$ reverse shock. The knot destruction time of $\sim 50$ years
is consistent with observed lifetimes of the optical fast moving knots
and quasi-stationary flocculi
\citep{thorstensen01,sutherland95,vandenbergh85,kamper76}. However the
X-ray knots of interest here are of similar temperature to their
surroundings (i.e., they have not cooled to optically emitting
temperatures), and their apparent survival for $\sim 200 - 300$ years
(for models in Table 1) places an upper limit on their density with
respect to the surrounding plasma of around a factor of 3 \citep[see
Table 3 of][]{klein94}. These authors did not consider underdense
knots, but the simulations of \citet{blondin01} suggest that these
would be no less unstable.  The X-ray knots do not appear to coincide
with the fast moving optical knots (FMKs), which do appear to be
significantly overdense and are currently undergoing shock
interaction.  We believe the X-ray knots underwent reverse shock
passage, and perhaps interacted with secondary shocks, some time early
in the evolution of Cas A and are now expanding with the rest of the
remnant plasma, as in fact appears to be the case \citep{delaney03}.

X-ray knots of similar density to the surroundings are also much more
consistent with the apparent factor of only 3-4 increase in surface
brightness compared to the ambient plasma. Consider a knot with solar
abundance ratios of Si, S, Ar, Ca, etc relative to O compared to a
pure O knot. At temperatures around 1 keV the Si-Ca is predominantly
in He- and H-like charge states and emits about 30 times more power
per ion \citep[$\sim 15$ times more power per unit mass;][]{summers79}
than O, which is fully ionized and radiates only by thermal
bremsstrahlung.  A knot composed of 0.75 O and 0.25 Si-Ca by mass will
thus radiate nearly 5 times more energy than a pure O knot integrated
over photon energy. When considering just emission in the X-ray
bandpass, the knot with heavy elements will radiate even more compared
with the pure O knot by an extra factor $\sim\exp\left(E/k_{\rm
B}T_e\right) \simeq 2$, where $E$ is the lower limit of the
bandpass. This occurs because most of the O thermal bremsstrahlung
power is radiated outside the X-ray band of interest, but the line
emission from the heavier elements is not. A knot of dimension $\sim
10^{17}$ cm compared with background plasma extending over 1 pc in
depth needs to be around 30 times more radiative in order to be
visible as a distinct object, so a density enhancement of around 2,
giving 4 times more radiation, is all that is required.  The knots
fitted in Table 1 all have Si/O abundance ratios typically 0.5$-$1 times
the solar value \citep[relative to the most recent value of the solar
O abundance,] []{allende01,holweger01}, and also usually have small
amounts of Fe which we have not considered here. The knot visibility
for minimal density enhancement becomes even stronger if the
surrounding plasma is in part H/He/N dominated composition, as would
be expected for circumstellar medium shocked by the blast wave or
ejecta shocked very early by the reverse shock.

\subsection{Observations and Data Analysis}
For our spectral analysis of ejecta knots in Cas A, we use the 50 ksec
observation with the Chandra Advanced CCD Imaging Spectrometer from
January 2000. The observation uses the backside-illuminated CCD S3 in
GRADED mode, wherein each CCD event is classified by its distribution
over the detector pixels before telemetry from the spacecraft.  We
have considered three radial series of knots to the east, northeast at
the base of the ejecta jet, and north-northwest, as shown in Figure 1.
The knots have angular diameters of a few to several arcseconds, and
are numbered starting closest to the center and progressing radially
outward, except for the base of the ejecta jet, which is shown in more
detail in Figure 2.  For each knot, we typically extract a spectrum
with several thousand counts prior to background subtraction.

The spectra are accumulated in pulse-invariant channels after
correction for the spatial variation of the gain across the detector,
but because the data were obtained in GRADED mode, the full removal of
the effects of charge transfer inefficiency was not possible.
Appropriate spatially weighted detector response functions and
effective area functions were constructed for each spectrum.  The
background was taken from several regions off the source.  For knots
which generally have high surface brightness relative to the
background (i.e., most of the knots in this sample), the spectral fits
are not highly sensitive to the background subtraction.  These data
were obtained relatively early in the Chandra mission (198 days after
launch) so the effect of the buildup of contaminants on the detector
is relatively small, but we include a component in the spectral model
to account for this excess absorption (ACISABS in XSPEC).

For each knot spectrum, we have fitted simple models for a single
temperature plasma with a single ionization age.  The spectral models
used do not include emission from the element Ar, so a narrow Gaussian
line is included in the model to represent the Ar He$\alpha$ blend
near 3.1 keV.  In all cases, the extremely strong emission lines
indicate that the emission is dominated by the ejecta, but an
additional continuum component is also required (as first noted by
Hughes et al. 2000).

We have considered two sets of assumptions for the source of this
continuum: the elements H, He, C, and N in their solar abundance
ratios, and an O-rich ejecta plasma in which H, He, C, and N, and Ne
are absent and O provides the continuum.  We favor the O continuum
fits as they are more straightforwardly compared to the model
calculations described above and are a more plausible description of
the composition of the ejecta (after Vink et al. 1996).  Both sets of
fits give reasonably comparable fits, although in specific cases,
either might at times give a better value fot the fit statistic than
the other.  We use the O continuum fit results, noting that the values
of the fitted temperature and ionization age are typically affected by
less than 15\% and 40\% respectively if the solar composition
continuum is used instead, though the effect is larger in a few
cases. A comparison of the fits for knots 4 in the E and NNW are shown
in Figures \ref{figE4} and \ref{figNNW4}, for both O-rich and
solar compositions.

Because of residual uncertainties in the gain of the detector (e.g.,
from charge transfer inefficiency), combined with the known spatially
dependent radial velocities of the X-ray emitting gas (e.g., Markert
et al. 1983), we have allowed the redshift of the model to be freely
fitted; likewise a Gaussian smoothing function whose width scales with
energy was also included in the model.

Table 1 shows the results of the spectral fits for all three radial
series, giving the temperatures, ionization ages, and abundances of
the elements Si and Fe (by number relative to O).  In most cases, we
obtain reasonable values of the fit statistic ($\chi^2/dof \sim$1.5).
Poorer fits (e.g., in the NNW and for the outer knots in the E)
generally arise from the presence of Fe K emission in the spectrum
that is not well-fitted by the model.  The fits are driven strongly by
the prominent Si line emission, but knots that are very rich in Fe
relative to Si are described by different parameters than the knots
with strong Si emission, specifically in having a higher ionization
age.  For knots that show a mixture of these characteristics, fully
successful models would have to include components to describe Si and
Fe separately.

In principle, the blast wave also contributes to each of the knot
spectra.  To verify that it is valid to neglect the blast wave
contribution in examining these compact regions, we have compared the
fit result for the well-isolated knot NNW1 using the standard
off-source background compared to a local background region
surrounding the knot.  The local background should incorporate the
blast wave contribution so that using this background spectrum
effectively subtracts the blast wave contribution at the knot.  We
found no significant difference in the temperature (kT =
0.93$^{+0.08}_{-0.09}$ keV compared to kT = 1.07$^{+0.12}_{-0.09}$ keV
with the off source background), and a poorly constrained ionization
age that is consistent with the value previously obtained.

\subsection{Ejecta Profile Models}
Our methods for computing the ejecta model follow from calculations in
\citet{laming01b}, \citet{laming02} and \citet{laming03}, and are
summarized in the two appendices to this paper. We take an analytic
approximation to the hydrodynamics for supernova ejecta expanding into
a remnant stellar wind from \citet{truelove99}, specified more
thoroughly in Appendix A. Within this framework we calculate the
time dependent ionization balance for the elemental composition of the
ejecta, the electron and ion temperatures allowing only for
equilibration between species by Coulomb collisions, and radiative and
adiabatic expansion power losses, as outlined in Appendix B.

We have calculated two series of ejecta models, summarized in Tables
\ref{tbl-1} and \ref{tbl-2}. Figures \ref{fig3} and \ref{fig4} show
the locus of electron temperature, $T_e$, against ionization age,
$n_et$, for these sets of models. We concentrate on these parameters
since they are readily determined from fits to the data, and the
curves of $T_e$ against $n_et$ are, for reasons discussed further
below, surprisingly robust against various changes to the model, being
mainly dependent on the ejecta density profile specified through the
envelope power law $n$. Both sets of models have an ejecta mass of
$M_{ej}=2M_{\sun}$. The first has $E_{51}=3\times 10^{51}$ ergs
explosion energy and an ambient density at the current blast wave
position in the range 3.1$-$3.6 H atoms cm$^{-3}$, depending on the
value taken for the blast wave radius (i.e. $\rho r_b^2=21$ H atoms
cm$^{-3}$ pc$^2$). These values are chosen to match the blast wave
velocity and radius measurements of \citet{delaney03} as closely as
possible for values of $n\simeq 9$ (in the middle of our range). The
second set has the same ejecta mass but an explosion energy of
$2\times 10^{51}$ ergs and circumstellar density reduced from before
by the same factor $2/3$. These models may be scaled keeping
$E_{51}\propto M_{ej} \propto \rho r_b^2$ which leaves the blast wave
velocity and radius invariant. The first set of models is chosen to
match the blast wave velocity and radius found by \citet{delaney03},
but implies a mass of shocked circumstellar material of close to 17
$M_{\sun}$, significantly larger than the $\sim 9 M_{\sun}$ found by
\citet{willingale02}.  The emission measure determined from thermal
bremsstrahlung using the BeppoSAX data studied in
\citet{laming01a,laming01b} interpreted as coming from an H-He
dominated plasma is consistent with masses of shocked circumstellar
plasma in the range 6.7$-$ 10 $M_{\sun}$ for densities ahead of the
blast at its current location of 3 $-$ 2 H atoms cm$^{-3}$ (or
equivalent mass) respectively. These densities combined with the blast
wave radius give masses of 11 $-$ 16.8 $M_{\sun}$ from simple
geometrical considerations for densities of 2 - 3 H atoms cm$^{-3}$.
It is clear that more consistent results derive from the lower
circumstellar density, but this gives a radius for the blast wave that
is slightly too small. We suspect the Cas A may have undergone an
initial period of expansion into a tenuous Wolf-Rayet wind before hitting the
much denser red supergiant wind into which it is now expanding. We
discuss this further in section 3.4.

In Tables \ref{tbl-1} and \ref{tbl-2} we give the predicted blast wave velocity, radius and
expansion parameter $\eta$, the reverse shock radius and the ratio of the shock radii.
This ratio is measured to vary between 1.52 and 1.73 with lower values being found in the
west limb of Cas A \citep{gotthelf01}. From the models we can see that this immediately
implies $n\simeq 9$ on the west limb decreasing to 6-8 elsewhere. The following table
entries give dynamical parameters from the shock solutions, and the final column gives
the mass of gas that may cool to optical emitting temperatures ($\sim 10^4$ K) within
320 years of explosion. This mass increases with $n$, being zero for the $n < 7$ models.
The initial inference of highest $n$ on the west limb is appealing since this is where
the strongest radio and non-thermal X-ray emission is located coincident with the position
of the contact discontinuity,
\citep{bleeker01,gotthelf01} implying that this is the
location of strongest magnetic field. The high mass of radiatively cooled and thus
very dense gas in this region will lead to stronger Rayleigh-Taylor instability at the
contact discontinuity here than elsewhere, and hence a higher magnetic field. In what
follows we will develop the idea of varying ejecta envelope power law with position in
the remnant and try to quantify the degree of anisotropy by reference to our ejecta models.

We argued above that only a modest degree of overdensity may exist in
the X-ray knots, in order for them to survive against hydrodynamic
instabilities following reverse shock passage. Here we demonstrate
that the clumping of ejecta by a factor up to around 3 does not
greatly affect the dependence of $T_e$ on $n_et$. Consider ejecta with
a reference value of $n_et$ at the current epoch, which encountered
the reverse shock at time $t_0$. If this were overdense by a factor
$\alpha $, say, then it would have passed through the reverse shock at
a time $t_c=\alpha t_0$, since the electron density varies as
$n_e\propto 1/t^2$ and $\int _{t_0}^{t^{\prime}}n_edt\propto 1/t_0$
for $t_0 << t^{\prime}$.  The clumped ejecta are shocked to a
temperature lower by $1/\alpha $ than the unclumped ejecta, but
undergoes less cooling by adiabatic expansion.  In ballistic expansion
temperatures are reduced by adiabatic expansion by a factor
\begin{equation}
T^{\prime}\left(t^{\prime}\right) \simeq T\left(t_0\right)
\exp\left(-{4\over 3}\int _{t_0}^{t^{\prime}} {v\over r}dt\right)
\simeq T\left(t_0\right)\exp\left(-{4\over 3}\int _{t_0}^{t^{\prime}}{dt\over t}\right)\simeq
T\left(t_0\right)\left(t_0\over t^{\prime}\right)^{4/3}.
\end{equation}
Hence clumped ejecta shocked at $\alpha t_0$ will have a final
temperature higher than that for unclumped ejecta by a factor $\alpha
^{-1}\times\alpha ^{4/3}=\alpha ^{1/3}$. In Sedov-Taylor expansion
$r\propto t^{2/3}$, the final exponent 4/3 become 8/9, and clumped
ejecta has a final temperature lower than that for unclumped ejecta by
a factor $\alpha ^{-1/9}$.  Additionally, if $T_i>>T_e$ so that
$T_e\propto \left(T_in_et\right)^{2/5}$, \citep{laming01a}, the
temperature difference factor varies as $\alpha ^{2/15}\rightarrow
\alpha ^{-2/45}$, remembering that the fits are sensitive only to
$T_e$. Consequently only for large overdensities $\alpha$m will
significant changes in $T_e$ result. We have verified with numerical
calculations that these conclusions hold, using formulae given by
\citet{sgro75} for transmitted and reflected shock velocities.

It has been argued previously \citep{laming01a,laming01b} that the
non-thermal hard X-ray continuum of Cas A extending out to 100 keV is
bremsstrahlung from a population of electrons accelerated by secondary
shocks propagating within the shell. These shocks arise as the forward
or reverse shocks encounter density contrasts and split into
transmitted and reflected shocks. How justified are we in assuming
that the ejecta knots of interest encounter the reverse shock and then
expand self similarly with the rest of the plasma, undergoing no
further interaction with secondary shocks? Following the discussion
and analysis in \citet{vink02}, we believe that the electron
acceleration actually occurs in shocked circumstellar plasma in the
immediate vicinity of the contact discontinuity, since an instability
operating in colder shocked ejecta is much less likely to yield the
required electron energies. Thus it is the blast wave encountering
quasi-stationary flocculi that gives rise to the secondary shocks of
interest, not the reverse shock. The ejecta knots we study here do in
fact have Lagrangian mass coordinates that place them well back from
the contact discontinuity, i.e. there is a lot of mass through which a
secondary shock originating at the blast wave would have to propagate
through to reach them, and we therefore consider the possibility of
significant heating by secondary shocks unlikely.  The reverse shock
is currently propagating through even more tenuous plasma than the
forward shock, and again is unlikely to produce secondary shocks that
perturb the ejecta knots. In our view, the ejecta were subject to
various hydrodynamic instabilities following reverse shock passage and
interactions with secondary shocks for a time comparable to the time
taken for the reverse shock to propagate further into lower density
ejecta. What we now see as ``knots'' are those structures that were
able to survive against these instabilities, due to having little
density contrast with their surroundings.

\subsection{Discussion}
Figures \ref{fig3} and \ref{fig4} plot the locus of $T_e$ against
$n_et$ for the two sets of ejecta models, superposed with results from
those spectral fits for which the $\chi ^2$ per degree of freedom is
less than 2. In both cases the knots from the jet base on the NE limb
consistently indicate the lowest value for $n$. In each case NE 3, 4,
6, 7, 8, 9, and 10 clearly indicate $n\le 6$ or $n\le 7-9$, depending on
choice of model,  and NE 5 (not plotted) is also
consistent with this.  NE 1,2, and 11 suggest $n\ge 12$
or $n=6$ if coming from the envelope portion of the curve. We suspect
that this relatively large change in $n$, if real, over small apparent
distances is mainly due to projection effects.  Knots NE 1,2 and 11 are
either in front of or behind NE 3, 4, 6, 7, 8, 9, and 10, which all
come from the brightest part of this region. Comparing to
\citet{delaney03}, our knots are closest to their regions 2$-$5 from
which they measured forward shock velocities in the range 4361$-$6520 km
s$^{-1}$ and radii 2.08$-$2.43 pc.  The series of knots on the east limb
all indicate $n\simeq 9$ or higher, except for knots E1 and E2 which
are the two nearest the projected SNR center and are closer to
$n=5.5-6$. \citet{delaney03} do not take any measurements from this region
of the blast wave, the closest being their region 29 some distance to
the south.  The NNW series of knots pose more problems in fitting the
Fe K region, but of those for which adequate fits can be found, NNW 4
indicates $n=5.5-6$ and NNW 5 and 7 give $n=6-12$, depending on the
hydrodynamic model, with $n=6-7$ being preferred.  These knots are
closest to regions 13 and 14 in \citet{delaney03}. At this point we
give more attention to the model with $E_{51}=2$ and $\rho r_b^2 =
14$. Reasons for doing so are given above with reference to the
shocked circumstellar material and the geometry of the blast wave. We
also find that the lowest temperature knots for the higher energy
model are only consistent with ejecta envelope power laws $n=30 - 50$
which are much steeper than any realistic explosion model would
predict. Hydrogen rich ejecta give electron temperatures lower for a
given $n_et$ by a factor 2$-$3. More electrons per baryon means that the
equilibrated electron temperature will always be a factor of 2 lower
than for O. Additionally, the lower ion charge leads to slower
electron-ion equilibration before this point is reached. We consider
that these points reinforce our assumption of O-rich ejecta.

We would like to interpret the different ejecta envelope power laws as
due to asymmetries in the explosion.  The models of Tables \ref{tbl-1}
and \ref{tbl-2} give different ejecta core densities.  Varying
$E_{51}$ to keep the core density invariant yields
\begin{equation}
E_{51}\propto M_{ej}^{5/3}{\left(n-3\right)^{5/3}\over n^{2/3}\left(n-5\right)}.
\end{equation}
and we can see that the lower values of $n$ actually correspond to higher explosion
energy in that particular direction. An $n=6$ model corresponds to 66\% more energy than
$n=9$, and $n=5.5$ to a factor 2.6 more.
The variation still leads to blast wave radii predicted to vary between
2.35 and 2.62 pc, with lower $n$ giving larger radii.
This is similar to the range of blast wave radii found by
\citet{delaney03}, but is not completely consistent with the locations they measure,
and so for the time being we further adjust all models to give the same blast wave
radius. These new models are given in Tables \ref{tbl-5} and \ref{tbl-4}. The degree
of explosion asymmetry one would infer from the shallower ejecta profile is now slightly
reduced; $n=6$ corresponds to about 40\% more energy and $n=5.5$ to about a
factor of 2 more than $n=9$ for our preferred
hydrodynamic model. The curves of $T_e$ against $n_et$ for these new models for
$n=5.5$ and 6 are plotted as dotted lines on Figures \ref{fig3} and \ref{fig4}.
For higher $n$ the new curves are negligibly different to the previous ones. We can see
that for both hydrodynamic models the NE knots at the jet base now clearly suggest
$n=5.5$ and not $n=5.5$, and other conclusions remain the same.

The degree of energy asymmetry in the initial explosion (about a factor of 2 comparing
$n$ derived from knots at the jet base with that coming from the ratio of forward and
reverse shock radii) is at the
lower end of that coming from simulations.  Two basic mechanisms of
asymmetric core collapse explosions have been discussed in the
literature. \citet{fryer00} and \citet{fryer03} model the explosion of
a rotating star in two and three dimensions respectively. Compared to
non-rotating explosions, the rotating core appears to be stabilized
against convection, and the core bounce that seeds the neutrino-driven
convection is also weakened. Both these effects are reduced in the
polar regions, resulting in an asymmetric explosion which is stronger
in the polar regions. Ejecta velocities about a factor of two higher
in polar relative to equatorial regions are predicted, corresponding
to a factor of four difference in kinetic energy if the density is the
same in polar and equatorial regions. $^{56}$Fe is generally
synthesized as $^{56}$Ni along the axis. Additionally, the generic
problem of core-collapse simulation producing too much neutron-rich
matter appears to be slightly exacerbated, though \citet{fryer02}
caution that much of the input physics upon which this conclusion
depends is still uncertain. In contrast \citet{khokhlov99} model a jet
induced explosion. This mechanism is independent of details of
convection. A magnetorotational instability in the collapsing core
\citep[recently studied in more detail by][]{akiyama02} accelerates
two jets along the rotation axis.  These drive bowshocks ahead of them
that compress most of the remaining ejecta into an equatorial
torus. Similar anisotropies in ejecta velocities to the rotating case
above are found, though the anisotropic ejecta density distribution
causes the kinetic energies to be less anisotropic.

\citet{nagataki98} postulate similar degrees of asymmetry in their
models to explain high yields of $^{44}$Ti produced in $\alpha$-rich
freeze out in the polar regions. Their model A1, with similar
asymmetry to models discussed above, produces $1.8\times 10^{-4}
M_{\sun}$ of $^{44}$Ti and $0.06 M_{\sun}$ of $^{56}$Ni for a suitable
choice of mass cut. The $^{44}$Ti mass is similar to that inferred for
Cas~A \citep{vink01,vink02}. The mass of $^{56}$Ni produced is less
reliably estimated from observations.  In a companion paper
\citep{hwang03} we do indeed find evidence of $^{44}$Ti production in
$\alpha$-rich freeze out, not in the jet region but further out along
the east limb series of knots.  Indeed this, the general distribution
of heavy elements in a torus around the ``jet'' axis, and the
existence of the ``jet'' itself in Cas A do resemble the model
discussed by \citet{khokhlov99}, though considerably more detailed
analysis would be required to confirm anything more quantitative.

The progenitor mass may now be estimated from the parameters of our models. Most of the
stellar wind is expelled during the red supergiant phase, and it is into this wind that
we believe Cas A is now expanding. A massive star spends $\sim 2\times 10^5$ years in this
stage \citep{garcia96}, roughly independent of stellar mass. The mass loss rate in this
phase is given by
\begin{equation}
{dM\over dt} = 4\pi\rho r^2v_w = 3\times 10^{-5}\left(\rho r_b^2\over {\rm 1~H~atom~cm}^{-3}
{\rm pc}^2\right)\left(v_w\over 100~{\rm km~s}^{-1}\right) M_{\sun}~{\rm year}^{-1},
\end{equation}
where $v_w$ is the stellar wind speed and our favored value for $\rho
r_b^2=14$.  We estimate $v_w$ from the speeds of the quasi-stationary
flocculi of Cas A. In general, they are observed to have proper
motions ranging up to around 500 km s$^{-1}$ \citep{vandenbergh85},
consistent with clumps a factor of $\sim 10^2 - 10^3$ more dense than
the ambient plasma having undergone some acceleration during the
passage of the blast wave. The lowest observed velocity is 20 km
s$^{-1}$ for QSF 10.  Taking this value in equation (3) we estimate a
total mass loss during the red supergiant phase of 17 $M_{\sun}$. The
iterated blast wave speeds and radii in Tables 4 and 5 are
respectively slightly higher and lower than the average values in
\citet{delaney03}, with our second set of models for lower $\rho
r_b^2$ being more discrepant. This might suggest an early stage of
expansion into a much more tenuous stellar wind from a Wolf-Rayet
progenitor, before the blast wave encounters the much more dense red
supergiant wind, similar to a model for Cas A proposed by
\citet{borkowski96}.  In the case of Cas~A such a fast tenuous wind
can not have existed for very long before explosion because one would
then lose the radiative instability of shocked ejecta and the
accompanying optical emission.  The mass lost during this period is
probably negligible. To the estimate above we should then add the
observed mass of ejecta and compact objects of about 3$-$4 $M_{\sun}$,
and the mass loss during the main sequence evolution to get a total
progenitor mass in the range 20-25 $M_{\sun}$, at the lower end of
Wolf-Rayet progenitor masses inferred by \citet{massey01}.  The
assumed value for the wind speed of 20 km s$^{-1}$ is significantly
lower than is usually assumed for red supergiants
\citep[e.g.][]{garcia96,lamers02} where speeds around 100 km s$^{-1}$
seem more plausible. This would increase our mass estimate to close to
100 $M_{\sun}$. This seems to us unlikely since the progenitor would
then spend much longer in the Wolf-Rayet phase before explosion
\citep{woosley93}, so much so that the Cas A blast wave should still
be moving through Wolf-Rayet wind rather than red supergiant wind,
with the result that no thermal instability should be present in the
ejecta, as discussed above. The Wolf-Rayet phase is much shorter for
less massive stars, again arguing for a progenitor at the lower end of
observed progenitor masses for Wolf-Rayet stars.  These masses are
consistent with those estimated from the initial $^{44}$Ti mass
inferred from $\gamma $ ray observations
\citep{iyudin94,iyudin97,vink01,vink02} compared with explosion
calculations \citep[e.g.][]{timmes96}. \citet{rothschild03} provide a
more complete discussion of this point.

\section{Conclusions}
This paper has presented an initial analysis of the X-ray ejecta knot
spectra from Cassiopeia A. Although considerably more labor intensive
and detailed than any previous analysis of imaging spectroscopy of
supernova remnants known to us, we believe that we are close to being
able to infer some of the really fundamental aspects of this object.
Rather than attempting to analyze and model the entire spectrum of the
remnant, we have isolated spectra from ejecta knots whose properties
we think we understand, which when combined with 1-D hydrodynamical
models allow us to infer ejecta envelope density profile power laws
and hence explosion energies in different directions in the remnant.
The fundamental discriminant in this work is the variation of the
reverse shock velocity in the models with different ejecta density
profiles. The faster reverse shock in shallower profiles leads to
higher ejecta temperatures and a larger separation between the forward
and reverse shocks. These considerations lead to conclusions
concerning the energy asymmetry of the explosion, which we find to be
around a factor of 2 larger in polar regions than at equatorial regions outside
the so-called jet. This is at the lower range of energy asymmetries
coming from existing core-collapse explosion simulations, and on this
basis alone we are not yet able to distinguish between rotating
convection driven explosions or the jet induced explosions. Aside from
the demonstration of new data analysis and interpretation techniques,
the one further important conclusion we wish to emphasize is that only
with significantly deeper Chandra observations of Cassiopeia A and
other supernova remnants will the full potential of these methods be
realized.

\acknowledgements
We wish to thank Larry Rudnick and Tracey Delaney for communication of their results
prior to publication, and particularly to Tracey Delaney for a careful check of some
of the material in Appendix A. JML was supported by basic research funds of the Office
of Naval Research.

\appendix
\section{Hydrodynamics for $s=2$ Supernova Remnants}
Here we summarize the equations governing the evolution of the forward and reverse shock
velocities and radii, following \citet{truelove99}. These authors concentrated on the
$s=0$ uniform density circumstellar medium case, so here we outline the extension of their
results to $s=2$, as implemented by us. The ejecta density profile is taken to
be a constant density core with an envelope obeying $\rho\left(r\right)\propto r^{-n}$ where
$n>5$. The circumstellar medium density profile is $\rho\left(r\right)\propto r^{-s}$ where
$s<3$. For a uniform density ambient medium $s=0$, and for a steady stellar wind, the case
we will take for Cas A, $s=2$. We work in similar units to \citet{truelove99};
\begin{eqnarray}
t_0=&423M_{ej}^{5/6}E_{51}^{-1/2}\rho ^{-1/3} {\rm years}\\
x_0=&3.07M_{ej}^{1/3}\rho ^{-1/3} {\rm pc}
\end{eqnarray}
for $s=0$ and
\begin{eqnarray}
t_0=&5633M_{ej}^{3/2}E_{51}^{-1/2}\left(\rho R_b^2\right)^{-1} {\rm years}\\
x_0=&40.74M_{ej}\left(\rho R_b^2\right)^{-1} {\rm pc}
\end{eqnarray}
for $s=2$, where $M_{ej}$ is the ejecta mass in solar masses, $E_{51}$ is the explosion
energy in units of $10^{51}$ ergs, $\rho $ is the circumstellar density at the blast wave
in hydrogen atoms (or equivalent mass) per cc and $R_b$ is the blast wave radius in pc.
The unit of time $t_0$ is related to the so-called Sedov-Taylor time in \citet{mckee95}
by $t_{ST}=t_0/2.024$. The unit of distance is similarly related by a factor
1.377 to their Sedov-Taylor distance $x_{ST}$. These fiducial quantities apply to the
case of uniform density ejecta expanding into a uniform density ambient medium.
For $t<t_{ST}$ the remnant is in the ejecta dominated phase where
the mass of the ejecta is dominant, and
for $t> t_{ST}$ the Sedov-Taylor phase where the swept up circumstellar mass dominates the
remnant evolution.

The ejecta dominated phase is further divided into an initial period while the
reverse shock is propagating through the ejecta envelope, and a later period once it
reaches the core.
During the initial envelope phase of evolution the
blast wave radius is given by
\citep[this is the generalization of equation (75 of][for arbitrary $s$]{truelove99}
\begin{equation}\label{eqnA1}
R_b=\left\{{l_{ED}^{\left(n-2\right)}\over\phi _{ED}}{3\left(3-s\right)^2\over
4\pi\left(n-3\right)n}\left[{10\over 3}{\left(n-5\right)\over\left(n-3\right)}
\right]^{\left(n-3\right)/2}\right\}^{1/\left(n-s\right)} t^{\left(n-3\right)/
\left(n-s\right)}.
\end{equation}
Here $l_{ED}$ is the ratio of the radii of the forward shock to
the reverse shock, known as the ``lead factor'', so that
$R_r=R_b/l_{ED}$, and we take $l_{ED}=1.19+8/n^2$. The ratio of
the pressures behind the reverse and forward shocks is given by
$\phi _{ED}=0.39-0.6\exp\left(-n/4\right)$. The values for these
parameters (both for $s=2$) are fitted to the tabulations given in
\citet{chevalier82}. The blast wave velocity is given by
\begin{equation}
v_b=\left(n-3\over n-s\right){R_b\over t}
\end{equation}
and the reverse shock velocity by
\begin{equation}
v_r={R_r\over t}-{dR_r\over dt} = \left(3-s\over n-3\right){v_b\over l_{ED}}.
\end{equation}

The blast wave radius during the phase when the reverse shock is propagating through
the ejecta core is given by \citep[from equation 45 of][]{truelove99}
\begin{equation}\label{eqnA4}
R_b^{\left(3-s\right)/2} +{3-s\over 3}\sqrt{l_{ED}f_0\over\phi _{ED}}
\left(R_b\over v_{ej}tl_{ED}\right)^{3/2}=\left(3-s\right)\sqrt{l_{ED}\over\phi
_{ED}} \left\{w_{core}^{3/2}f_0^{1/2}{n\over 3\left(n-3\right)} - {f_n^{1/2}\over n-3}
\right\},
\end{equation}
where $w_{core}=v_{core}/v_{ej}$ and
$v_{core}=\sqrt{10/3}\sqrt{\left(n-5\right)/\left(n-3\right)}$ is the
ejecta velocity at the core-envelope boundary and $v_{ej}$ is the ejecta velocity
at the outer edge of the envelope. In deriving equation A5 we have put
$w_{core}\rightarrow 1$. The ejecta structure function
$f\left(w\right)=f_0$ for $0\le w\le w_{core}$ and $f\left(w\right)=f_n/w^n$ for
$w_{core}\le w\le 1$. From continuity at the core-envelope boundary
$f_0=f_n/w_{core}^n\rightarrow 3/4\pi$ as $w_{core}\rightarrow 1$. The time $t_{core}$
where the reverse shock hits the ejecta core is determined from equation A5 by setting
$R_b=l_{ED}v_{core}t_{core}$, $w_{core}\rightarrow 1$ so that $v_{core}=v_{ej}$ and
solving for $t_{core}$ with the result
\begin{equation}
t_{core}={R_b\over l_{ED}v_{core}}=
\left\{{1\over l_{ED}^{\left(2-s\right)}\phi _{ED}}{3\left(3-s\right)^2\over
4\pi\left(n-3\right)n}\right\}^{1/\left(3-s\right)}
\left[{3\over 10}{\left(n-3\right)\over\left(n-5\right)}
\right]^{1/2}.
\end{equation}
This generalizes equation 79 of \citet{truelove99}. Equation \ref{eqnA4} is
quite simple to work with for $s=0$, but for $s=2$ it requires the solution of a cubic
equation in $R_b^{1/2}$, and for arbitrary $s$ can be even more complicated.
We adopt a simpler procedure of extending the blast wave envelope solution
into the core phase and matching it to the offset power law solution that is appropriate
in the Sedov-Taylor limit,
\begin{equation}
v_b={n-3\over n-s}{R_b\over t}={2\xi _s^{1/2}\over 5-s}R_b^{\left(s-3\right)/2}
\end{equation}
where $\xi = \left(5-s\right)\left(10-3s\right)/8\pi $.
The time at which these two solutions connect is derived by eliminating $v_b$ between
equations A6 and A9, and then substituting the envelope
solution for $R_b$, equation \ref{eqnA1} into the resulting expression for $t$ and solving
for $t$ with solution
\begin{equation}
t_{conn}=\left({n-3\over n-s}\sqrt{2\pi {5-s\over10-3s}}\right)^{2\left(n-s\right)/
\left(5-n\right)\left(3-s\right)}\left(v_{ej}l_{ED}\right)^{\left(n-s\right)\left(5-s\right)
/\left(5-n\right)\left(3-s\right)}t_{core}^{\left(5-s\right)/\left(5-n\right)}.
\end{equation}
The complete expression for the forward shock radius is
\begin{equation}
R_b=\left\{\left[\left\{l_{ED}^{n-2}3\left(3-s\right)^2\over
\phi _{ED}4\pi n\left(n-3\right)\right\}
\left(t_{conn}v_{core}\right)^{n-3}\right]^{\left(5-s\right)/2\left(n-s\right)}
+\sqrt{\left(5-s\right)\left(10-3s\right)\over 8\pi}
\left(t-t_{conn}\right)\right\}^{2/\left(5-s\right)}
\end{equation}
with the velocity given by equation A10.

The reverse shock trajectory through the ejecta core and into the Sedov-Taylor
phase cannot in general be specified without recourse to numerical calculations.
\citet{truelove99} show that for $s=0$ SNRs the reverse shock velocity is approximately
proportional to time and so introduce a constant acceleration parameterization, with
the value of the acceleration being determined numerically. In general the acceleration
turns out to be quite small, and the reverse shock velocity is almost constant.
By the same methods it
can be shown that for $s=2$ SNRs the reverse shock velocity varies as $\sqrt{t}$, and following
reverse shock propagation into the ejecta core, we expect it to accelerate much
less than in the $s=0$ case, and so during the core-Sedov-Taylor phase we hold
$v_r$ constant at the value given by equation A7 upon entry into the ejecta core. The
reverse shock radius in this phase is given by
\begin{equation}
R_r=\left\{{R_b\left(t=t_{core}\right)\over l_{ED}t_{core}}-v_r\ln\left(t/t_{core}
\right)\right\}t.
\end{equation}

To summarize, the blast wave radius and velocity are given by equations A45 and A5
for $t < t_{conn}$, given by equation A11, and are given by equations A10 and A12
for $t > t_{conn}$. The reverse shock radius and velocity (with respect to the
otherwise freely expanding ejecta) are given by equations A5 (divided by $l_{ED}$)
and A7 for $t < t_{core}$, with $t_{core}$ given by equation A9. For $t > t_{core}$
the reverse shock velocity is held constant and the radius is given by
equation A13.

\citet{truelove99} give no guidance on the motion of the contact discontinuity. We use
the results of \citet{chevalier82} for the envelope phase, and during core propagation
we assume that the contact discontinuity expands with 0.75 of the forward shock velocity.
Detailed calculations indicate
that the forward shock-contact discontinuity separation increases once the reverse
shock has reached its maximum radius \citep{wang01}. We estimate that this will occur
once Cas A is about 770 years old for the $n=6$ ejecta profile, the with projected time
increasing rapidly with increasing $n$. An inconsistency arises with this assumption for
the steeper models, in that the reverse shock radius can be {\em ahead} of the
contact discontinuity. We suspect that our estimate of the reverse shock velocity is
becoming inaccurate, and increase $v_r$ in these cases so that $R_r$ as given by
equation A13 is always less than $R_c=0.75R_b$. This modification produces essentially
no change in the spectroscopic parameters, $T_e$ and $n_et$ in which we are interested.
We have also tried several different analytic representations of the hydrodynamics and
find our curves of $T_e$ against $n_et$ to be insensitive to these differences.

\section{BLASPHEMER Simulations}
BLASPHEMER (BLASt Propagation in Highly EMitting EnviRonment) follows the time dependent
ionization balance and temperatures of a Lagrangian plasma parcel as it passes through
either the forward or reverse shock and then expands with the rest of the supernova remnant.
In this appendix we repeat and update some of the description in \citet{laming01b}.
Behind the reverse shock, the density $n_q$ of ions with charge $q$ is
given by
\begin{equation} {dn_q\over dt} =
n_e\left(C_{ion,q-1}n_{q-1}-C_{ion,q}n_q\right) +
n_e\left[\left(C_{rr,q+1} +C_{dr,q+1}\right)n_{q+1} -
\left(C_{rr,q}+ C_{dr,q}\right)n_q\right]
\label{eqn1}\end{equation}
where $C_{ion,q}, C_{rr,q}, C_{dr,q}$ are
the rates for electron impact ionization, radiative recombination and dielectronic
recombination respectively, out of the charge
state $q$. These rates are the same as those used in the recent ionization balance
calculations of \citet{mazzotta98}, using subroutines kindly supplied by
Dr P. Mazzotta (private communication 2000). The electron density $n_e$ is determined
from the condition that the plasma be electrically neutral. The ion and electron
temperatures, $T_i$ and $T_e$ are given by
\begin{equation} {dT_i\over dt}= -0.13n_e{\left(T_i-T_e\right)\over AT_e^{3/2}}
{\sum _qq^3n_q/\left(q+1\right)\over\sum _q n_q}
\end{equation} and
\begin{equation} {dT_e\over dt}= 0.13n_e{\left(T_i-T_e\right)\over AT_e^{3/2}}
{\sum _qq^2n_q/\left(q+1\right)\over\sum _q n_q}
-{T_e\over n_e}{dn_e\over dt} - {2\over 3n_ek_{\rm B}}{dQ\over dt}.
\end{equation}
Here $A$ is the atomic mass of the ions in the plasma. The last term $dQ/dT$
represents plasma energy losses due to ionization and radiation. Radiation losses
are taken from \citet{summers79}. At each time step $T_i$ and $T_e$ are modified
by a further factor $\exp\left(-4v_{ex}\Delta t/3r\right)$ and $n_e$ and the $n_q$ by
$\exp\left(-2v_{ex}\Delta t/r\right)$. Here $v_{ex}$ is the
expansion velocity, which is assumed constant for all ejecta, and is
given by the expansion velocity of the contact discontinuity discussed above.
The radial compression/decompression of shells of shocked ejecta is
treated by further modifying densities and temperatures by $v_{ex}/v_{ex}^{\prime}$
and $\left(v_{ex}/v_{ex}^{\prime}\right)^{2/3}$ respectively at each time step,
where $v_{ex}^{\prime}$ is $v_{ex}$ at the previous time step.
The plasma pressure is evolved according to
adiabatic expansion in the same way, and the densities and temperatures further
adjusted by $\left(P/P^{\prime}\right)^a$ where $a=0.6$ for densities and 0.4 for
temperatures. This accounts for the compression of plasma due to radiation and ionization
losses. The initial plasma pressure is $P$ and the plasma pressure after losses is
$P^{\prime }$, and the Lagrangian plasma element is recompressed adiabatically at each
time step to restore its pressure to the adiabatic value after losses. Except for
ejecta near the contact discontinuity, radiation losses are generally unimportant.

\citet{spitzer78} gives the timescale for an electron distribution to relax to a
Maxwellian as
\begin{equation}
t_{eq}\left(e,e\right)={3m_e^{1/2}\left(k_{\rm B}T_e\right)^{3/2}\over 8\sqrt{2\pi }
n_ee^4\ln\Lambda }
\end{equation}
where $\Lambda$ is the so-called plasma parameter, the ratio of largest to smallest
impact parameters for collisions. In supernova remnants $\ln\Lambda\simeq 40$. The
equilibration time for ions
$t_{eq}\left(i,i\right)=t_{eq}\left(e,e\right)\sqrt{m_i/m_e}/Z_i^4$, and that for
electron-ion equilibration is $t_{eq}\left(e,i\right)=t_{eq}\left(e,e\right)m_i/m_e/Z_i^2$
where $Z_i$ is the ion charge. Accordingly we write
\begin{equation}
{d\Delta T\over dt}=-0.13Z^2n_e{\Delta T\over AT_e^{3/2}}
\end{equation}
which is equation (2), with $\Delta T=T_i-T_e$. We consider a fully ionized gas with
$n_e=Zn_i$ and
\begin{equation}
{d\over dt}\left(n_iT_i+n_eT_e\right)=n_i{dT_i\over dt}+ n_e{dT_e\over dt}=0.
\end{equation}
Solving these equations yields
\begin{eqnarray}
{dT_e\over dt}=0.13{Z^2n_e\over Z+1}{T_i-T_e\over AT_e^{3/2}}\\
{dT_i\over dt}=-0.13{Z^3n_e\over Z+1}{T_i-T_e\over AT_e^{3/2}}.
\end{eqnarray}
In deriving equations (14) and (15) these expressions are averaged over
the ion charge states in the plasma, and the expression for $dT_e/dt$ is
modified by the inclusion of terms accounting for the change in electron
density due to ionization, $-\left(T_e/n_e\right)\left(dn_e/dt\right)_{ion}$,
and radiative and ionization losses, $-\left(2/3n_ek_{\rm B}\right)dQ/dt$.
Recombinations, which reduce the electron density do not result in an increase
in the electron temperature in low density plasmas, since the energy of the
recombined electron is radiated away, rather than being shared with the other
plasma electrons as would be the case for three-body recombination in dense plasmas.

\clearpage



\plotone{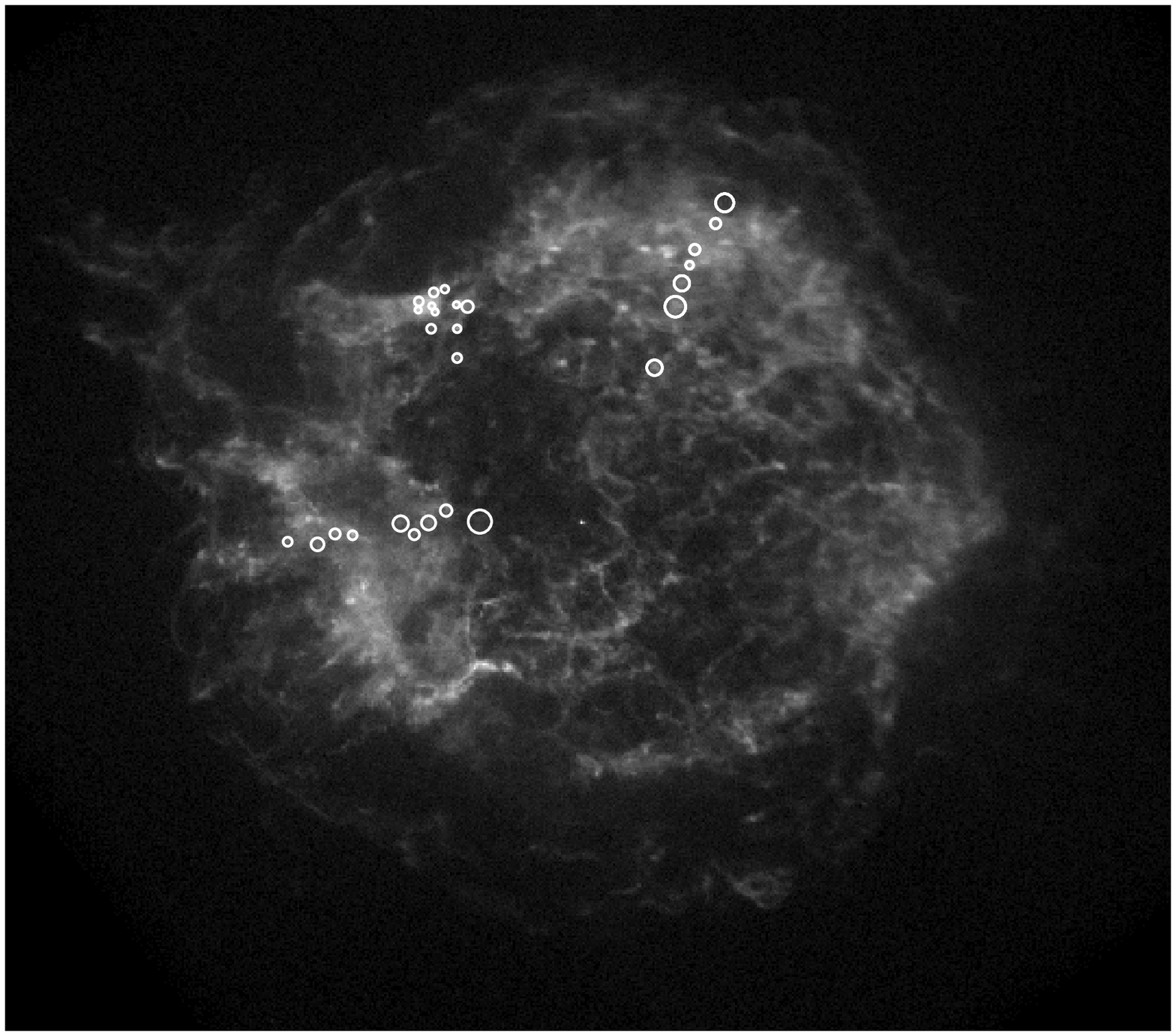}
\figcaption[f1.ps] {Locations of series of ``Si'' rich knots on northern limb.
East is to the left and north is up.
\label{fig1}}

\plotone{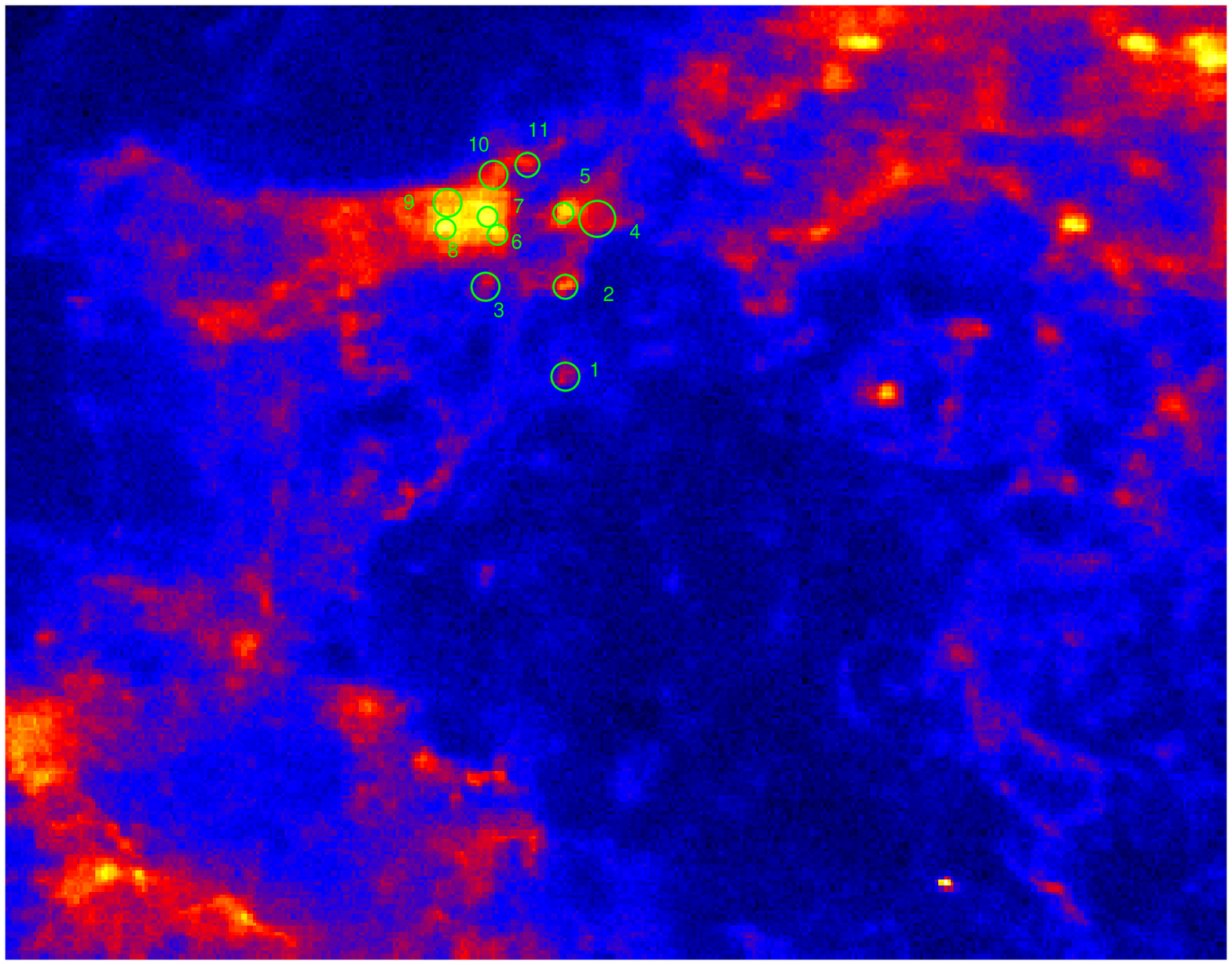}
\figcaption[f2.ps] {Detail of Cas A showing locations of ``Si'' rich knots
at the jet base in the NE region. East is to the left and north is up.\label{fig2}}

\begin{figure}
\begin{center}
\centerline{\hbox{\psfig{figure=f3a.ps,width=8cm,angle=-90}
\psfig{figure=f3b.ps,width=8cm,angle=-90}}}
\caption{Fits to spectrum of knot E4 with O dominated
composition (left) and H/He dominated composition (right).\label{figE4}}
\end{center}
\end{figure}

\begin{figure}
\begin{center}
\centerline{\hbox{\psfig{figure=f4a.ps,width=8cm,angle=-90}
\psfig{figure=f4b.ps,width=8cm,angle=-90}}}
\caption{Fits to spectrum of knot NNW4 with O dominated
composition (left) and H/He dominated composition (right).\label{figNNW4}}
\end{center}
\end{figure}

\plotone{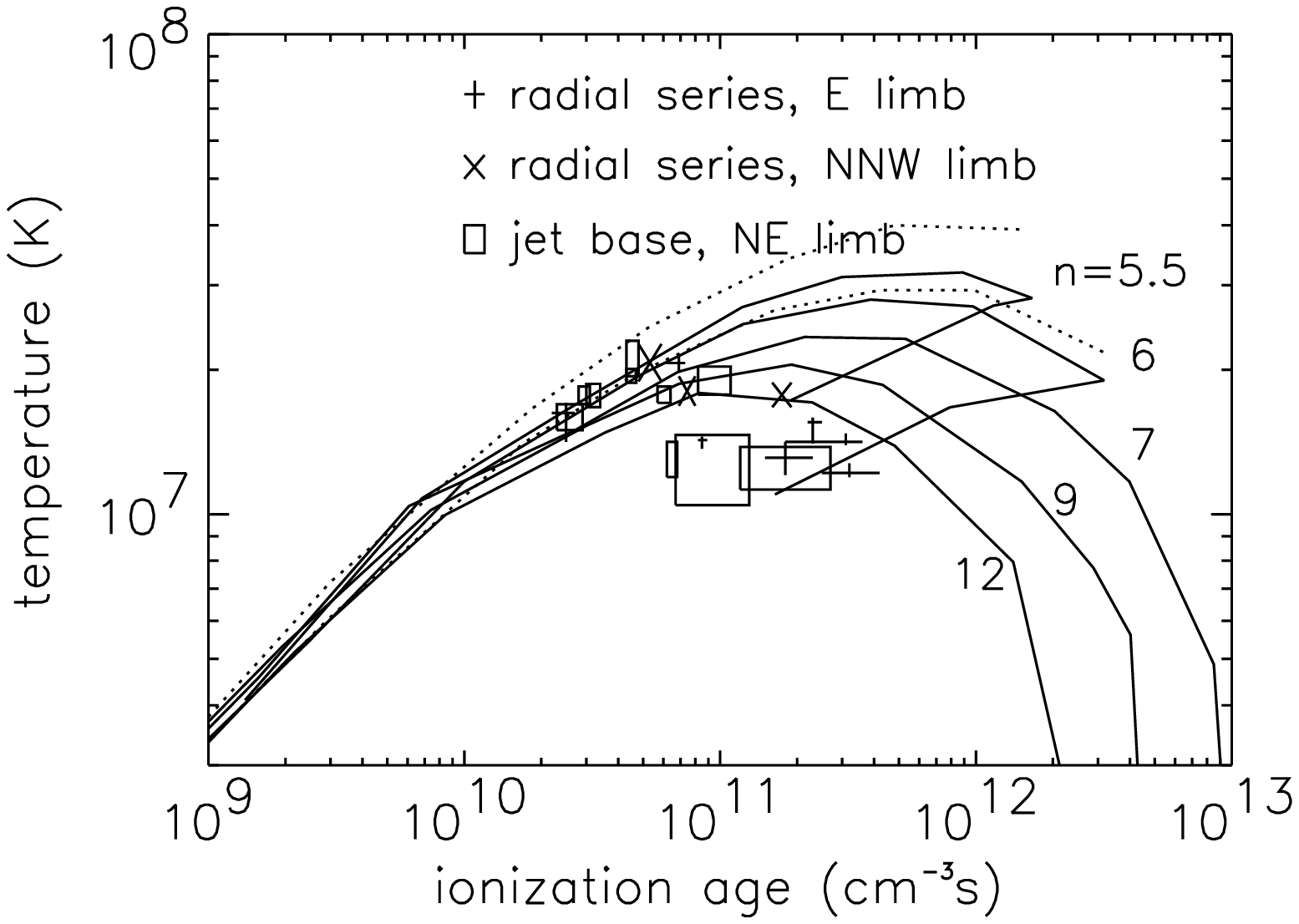}
\figcaption[f5.ps]{Plots of $T_e$ against $n_et$ for varying ejecta envelope
power laws. The solid lines give models with $M_{ej}=2M_{\sun}$, $E_{51}=3$ and
$\rho r_b^2=21$, for $n=5,5$, 6, 7, 9, and 12 from the top. The point at highest $n_et$
for $n=5.5$ and 6 corresponds to ejecta at the core-envelope boundary. For higher values
of $n$ this plasma undergoes thermal instability. The dotted lines indicate the
variation of $T_e$ with $n_et$ for the modified models in Table 4 for $n=5.5, 6$.
Higher values of $n$ change insignificantly in the modified models. Only fits with
$\chi ^2/$d.o.f. $< 2$ are plotted.\label{fig3}}

\plotone{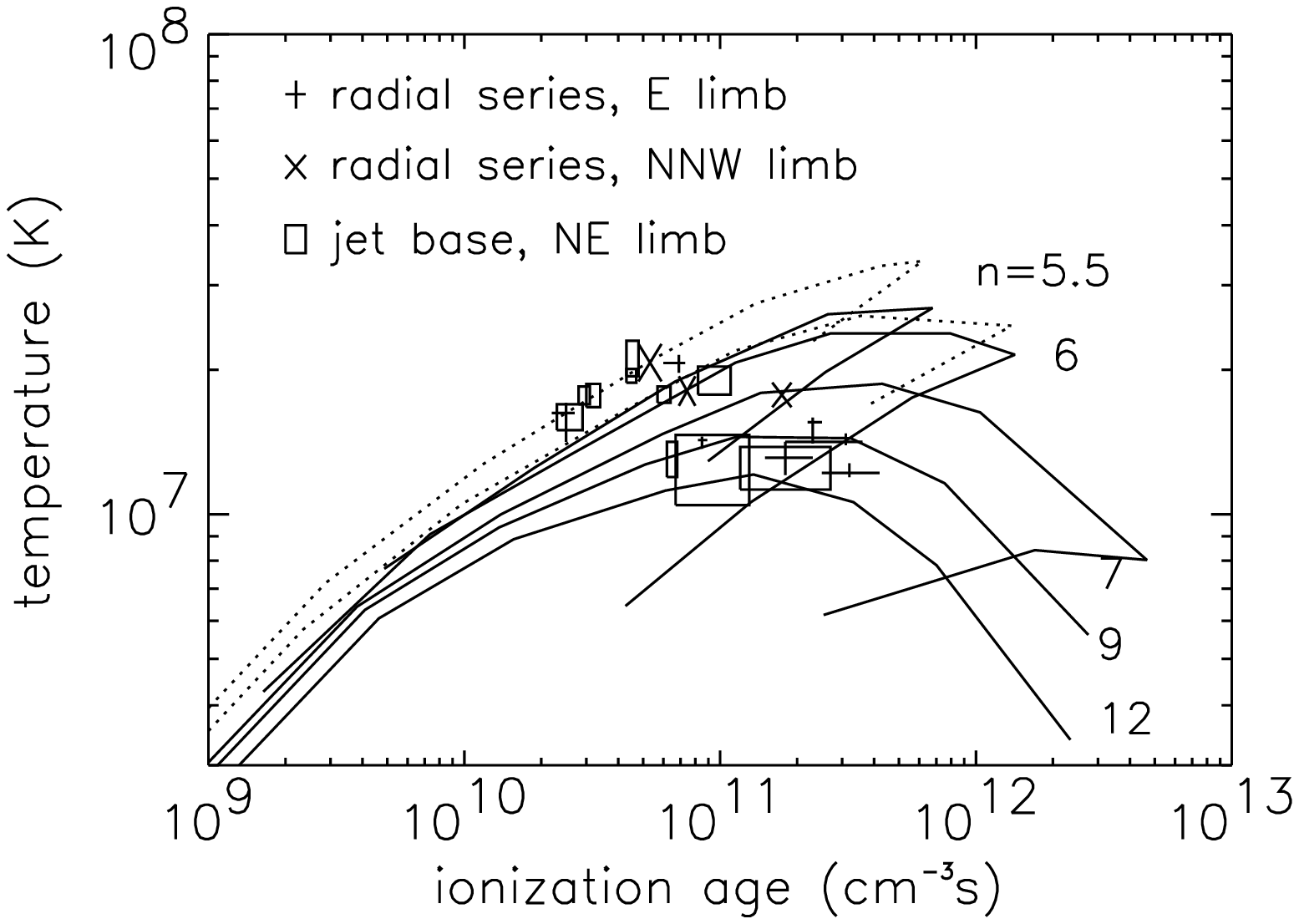}
\figcaption[f6.ps]{Plots of $T_e$ against $n_et$ for varying ejecta envelope
power laws. The solid lines give models with $M_{ej}=2M_{\sun}$, $E_{51}=2$ and
$\rho r_b^2=14$, for $n=5.5$, 6, 7, 9, and 12 from the top. The point at highest $n_et$
for $n=5.5$, 6, and 7 corresponds to ejecta at the core-envelope boundary.
For higher values
of $n$ this plasma undergoes thermal instability. The dotted lines indicate the
variation of $T_e$ with $n_et$ for the modified models in Table 5 for $n=5.5, 6$.
Higher values of $n$ change insignificantly in the modified models. Only fits with
$\chi ^2/$d.o.f. $< 2$ are plotted.\label{fig4}}

\clearpage
\begin{deluxetable}{cccccccccccc}
\tabletypesize{\scriptsize}
\tablecaption{Fits to O-Si rich Knots\label{tbl-3}}
\tablewidth{0pt}
\tablehead{
\colhead{knot\tablenotemark{a}} &
\colhead{$k_{\rm B}T_e$ (keV)\tablenotemark{b}} &
\colhead{$n_et$ (cm$^{-3}$s)} & \colhead{Si/O\tablenotemark{c}} &
\colhead{Fe/O} & \colhead{$n_{\rm H}$ ($10^{22}$ cm$^{-2}$)} &
\colhead{$\chi ^2$} & \colhead{$\chi ^2$/dof} & \colhead{no. dof} & \colhead{no. cts} \\
  }

\startdata
E1 & 1.78 &6.9e+10 &  0.08 &  0.05 &  1.60  &  233.1  &  1.46 &  160 & 9832 \\
          & 1.70-1.85 & 6.0e+10-7.3e+10 & 0.07-0.09 & 0.04-0.06 & 1.49-1.67 & & & & \\
E2 & 1.40 &2.5e+10 &  0.35 &  0.04 &  1.99  &  199.9  &  1.54 &  130 & 7976 \\
          & 1.22-1.47 & 2.2e+10-2.7e+10 & 0.31-0.37 & 0.03-0.04 & 1.96-2.05 & & & & \\
E3 & 1.05 &3.2e+11 &  0.28 &  0.17 &  1.22  &  236.7  &  1.75 &  135 & 11897 \\
          & 1.03-1.10 & 2.5e+11-4.2e+11 & 0.24-0.39 & 0.13-0.20 & 1.16-1.28 & & & & \\
E4 & 1.13 &1.8e+11 &  0.29 &  0.07 &  1.11  &  184.7  &  1.71 &  108 & 5994 \\
          & 1.04-1.22 & 1.5e+11-2.3e+11 & 0.27-0.35 & 0.06-0.10 & 1.03-1.24 & & & & \\
E5 & 1.34 &2.3e+11 &  0.29 &  0.12 &  0.99  &  236.9  &  1.74 &  136 & 10797 \\
          & 1.21-1.37 & 2.2e+11-2.5e+11 & 0.27-0.31 & 0.09-0.12 & 0.96-1.02 & & & & \\
E6 & 1.23 &8.5e+10 &  0.29 &  0.05 &  1.51  &  271.8  &  1.84 &  148 & 10452 \\
          & 1.18-1.25 & 8.2e+10-8.9e+10 & 0.28-0.34 & 0.04-0.06 & 1.49-1.58 & & & & \\
E7 & 0.91 &2.0e+11 &  0.28 &  0.04 &  1.2  &  332.8  &  2.47 &  135 & 9389 \\
E8 & 0.72 &4.2e+11 &  0.45 &  0.04 &  1.2  &  296.0  &  2.29 &  129 & 10316 \\
E9 & 1.22 &3.1e+11 &  0.28 &  0.13 &  1.20  &  163.6  &  1.56 &  105 & 4924 \\
          & 1.20-1.27 & 1.8e+11-3.6e+11 & 0.23-0.36 & 0.11-0.21 & 1.17-1.28 & & & & \\
\\
NNW1 & 1.42 &1.1e+11 &  0.29 &  0.12  &  1.3  &   335.8 &   2.05 &  164 & 15126 \\
NNW2 & 1.85 &1.2e+11 &  0.24 &  0.11  &  1.1  &   760.7 &   3.21 &  237 & 32844 \\
NNW3 & 1.75 &1.4e+11 &  0.21 &  0.09  &  1.0  &   548.1 &   3.11 &  176 & 16483 \\
NNW4 & 1.86 &5.1e+10 &  0.32 &  0.11  &  1.49  &   203.3 &   1.42 &  143 & 8725 \\
            & 1.63-1.95 & 4.8e+10-5.9e+10 & 0.29-0.34 & 0.10-0.12 & 1.45-1.54 & & & & \\
NNW5 & 1.49 &7.6e+10 &  0.24 &  0.08  &  1.22  &   242.9 &   1.48 &  164 & 12960 \\
            & 1.45-1.67 & 6.9e+10-8.0e+10 & 0.22-0.25 & 0.07-0.09 & 1.20-1.28 & & & & \\
NNW6 & 1.10 &3.3e+10 &  0.20 &  0.00  &  1.7  &   305.0 &   2.18 &  140 & 10584 \\
NNW7 & 1.54 &1.7e+11 &  0.18 &  0.04  &  0.84 &   210.9 &   1.70 &  124 & 6826 \\
            & 1.44-1.62 & 1.6e+11-1.9e+11 & 0.17-0.19 & 0.03-0.04 & 0.76-0.86 & & & & \\
\\
NE1  &1.11 & 1.6e+11 &  0.06 &  0.08 &  1.08 &   87.4  &  1.00 &   87 & 4124 \\
           & 0.97-1.19 & 1.2e+11-2.7e+11 & 0.05-0.08 & 0.06-0.10 & 0.95-1.24 & & & & \\
NE2  &1.12 & 6.5e+10 &  0.40 &  0.04 &  1.24 &  189.1  &  1.70 &  111 & 6009 \\
           & 1.03-1.22 & 6.2e+10-6.8e+10 & 0.38-0.52 & 0.03-0.04 & 1.10-1.29 & & & & \\
NE3  &1.54 & 3.2e+10 &  0.29 &  0.08 &  1.35 &  119.6  &  1.34 &   90 & 4155 \\
           & 1.44-1.61 & 3.0e+10-3.4e+10 & 0.27-0.31 & 0.07-0.09 & 1.32-1.37 & & & \\
NE4  &1.53 & 3.0e+10 &  0.60 &  0.04 &  1.29 &  266.8  &  1.89 &  141 & 10884 \\
           & 1.46-1.59 & 2.8e+10-3.1e+10 & 0.57-0.62 & 0.03-0.04 & 1.27-1.33 & & & & \\
NE5  &1.51 & 3.3e+10 &  0.44 &  0.03 &  1.4 &  284.1  &  2.47 &  115 & 6961 \\
NE6  &1.32 & 2.8e+10 &  0.34 &  0.00 &  1.61 &  191.4  &  1.48 &  129 & 7578 \\
           & 1.29-1.46 & 2.3e+10-2.9e+10 & 0.32-0.36 & 0.00-0.003 & 1.60-1.63 & & & & \\
NE7  &1.67 & 4.5e+10 &  0.25 &  0.06 &  1.54 &  280.4  &  1.76 &  159 & 10016 \\
           & 1.62-1.73 & 4.3e+10-4.7e+10 & 0.24-0.26 & 0.05-0.06 & 1.52-1.56 & & & & \\
NE8  &1.90 & 4.3e+10 &  0.26 &  0.04 &  1.52 &  248.0  &  1.53 &  162 & 9776 \\
           & 1.67-1.98 & 4.3e+10-4.8e+10 & 0.23-0.27 & 0.03-0.04 & 1.48-1.56 & & & & \\
NE9  &1.51 & 6.1e+10 &  0.21 &  0.05 &  1.31 &  308.2  &  1.78 &  173 & 14081 \\
           & 1.47-1.59 & 5.7e+10-6.4e+10 & 0.19-0.22 & 0.04-0.05 & 1.28-1.32 & & & & \\
NE10 &1.67 & 9.1e+10 &  0.06 &  0.09 &  1.31 &  221.9  &  1.66 &  134 & 8451 \\
           & 1.53-1.75 & 8.2e+10-1.1e+11 & 0.05-0.08 & 0.07-0.10 & 1.27-1.38 & & & & \\
NE11 &1.10 & 9.1e+10 &  0.12 &  0.17 &  1.32 &  162.7  &  1.75 &   93 & 4812 \\
           & 0.90-1.26 & 6.7e+10-1.3e+11 & 0.09-0.15 & 0.14-0.21 & 1.22-1.41 & & & & \\
\enddata

\tablenotetext{a}{Knots are numbered in the E and NNW regions starting with the
innermost and working out to the limb. The knot numbering in the NNE region is given
in Figure 2.}
\tablenotetext{b}{Electron temperature in keV.}
\tablenotetext{c}{Element abundance ratio by number relative to solar values of
\citet{anders89}. Note that these are superseded by \citet{grevesse98}, \citet{allende01},
and \citet{allende02}. In particular abundances relative to O increase by 1.75.}

\end{deluxetable}

\begin{deluxetable}{ccccccccccc}
\tabletypesize{\scriptsize}
\tablecaption{Cas A Ejecta Profile Models $M_{ej}=2M_{\sun}$, $E_{51}=3$,
$\rho r_b^2=21$ \label{tbl-1}}
\tablewidth{0pt}
\tablehead{
\colhead{$n$} &
\colhead{$v_b\left(320 {\rm yrs}\right)$} &
\colhead{$r_b\left(320 {\rm yrs}\right)$} & \colhead{$\eta $\tablenotemark{a}} &
\colhead{$r_r\left(320 {\rm yrs}\right)$} & \colhead{$r_b/r_r$} &
\colhead{$v_{core}$\tablenotemark{b}} &
\colhead{$t_{core}$\tablenotemark{c}} & \colhead{$t_{conn}$\tablenotemark{d}} &
\colhead{$t_{rad}$\tablenotemark{e}}  &
\colhead{$M_{rad}$\tablenotemark{f}}\\
 & \colhead{km s$^{-1}$}
 & \colhead{pc} &  & \colhead{pc} &  & \colhead{km s$^{-1}$}
 & \colhead{yrs} & \colhead{yrs}& \colhead{yrs}& \colhead{$M_{\sun}$}
  }

\startdata
5.5& 4044 & 1.85& 0.72& 0.92& 2.00& 7098& 39.1& 50586& - & 0\\
6 & 4942 & 2.15& 0.75& 0.97& 2.21& 9163& 21.5& 632& -& 0\\
7 & 5139 & 2.36& 0.71& 1.13& 2.08& 11223& 10.1& 120.2& 10.05-10.25& 0.012\\
8 & 5108 & 2.38& 0.70& 1.35& 1.77& 12294& 5.99& 72.4& 4.55-10.4& 0.41\\
9 & 5095 & 2.40& 0.70& 1.58& 1.52& 12959& 3.97& 55.0& 2.67-11.1& 0.61\\
10 & 5088& 2.41& 0.69& 1.79& 1.34& 13414& 2.84& 45.7& 1.76-11.4& 0.69\\
11 & 5083& 2.41& 0.69& 1.99& 1.21& 13745& 2.14& 39.8& 1.25-11.9& 0.75\\
12 &5079 & 2.41& 0.69& 2.17& 1.11& 13997& 1.67& 35.6& 0.93-12.5& 0.79\\
\enddata


\tablenotetext{a}{Forward shock expansion parameter.}
\tablenotetext{b}{Free expansion
velocity of ejecta core-envelope boundary.}
\tablenotetext{c}{Time
following explosion when reverse shock enters ejecta core.}
\tablenotetext{d}{Time when blast wave solutions are connected.}
\tablenotetext{e}{Time interval for which ejecta passing through
the reverse shock cools to optically emitting temperatures within
320 years.}
\tablenotetext{f}{Mass of gas that can cool to
optically emitting temperatures within 320 years of explosion.}

\end{deluxetable}

\begin{deluxetable}{ccccccccccc}
\tabletypesize{\scriptsize}
\tablecaption{Cas A Ejecta Profile Models $M_{ej}=2M_{\sun}$, $E_{51}=2$,
$\rho r_b^2=14$ \label{tbl-2}}
\tablewidth{0pt}
\tablehead{
\colhead{$n$} &
\colhead{$v_b\left(320 {\rm yrs}\right)$} &
\colhead{$r_b\left(320 {\rm yrs}\right)$} & \colhead{$\eta $\tablenotemark{a}} &
\colhead{$r_r\left(320 {\rm yrs}\right)$} & \colhead{$r_b/r_r$} &
\colhead{$v_{core}$\tablenotemark{b}} &
\colhead{$t_{core}$\tablenotemark{c}} & \colhead{$t_{conn}$\tablenotemark{d}} &
\colhead{$t_{rad}$\tablenotemark{e}}  &
\colhead{$M_{rad}$\tablenotemark{f}}\\
 & \colhead{km s$^{-1}$}
 & \colhead{pc} &  & \colhead{pc} &  & \colhead{km s$^{-1}$}
 & \colhead{yrs} & \colhead{yrs}& \colhead{yrs}& \colhead{$M_{\sun}$}
  }

\startdata
5.5& 3928 & 1.79& 0.72& 1.08& 1.66& 5795& 71.8& 92930& - & 0\\
6& 4698 & 2.04& 0.75& 1.17& 1.75& 7482& 39.5& 1162& - & 0\\
7& 5239& 2.27& 0.76& 1.29& 1.76& 9163& 18.6 & 221& - & 0\\
8& 5177& 2.32& 0.73& 1.44& 1.62& 10038& 11.0& 133& 9.25-15.5& 0.26\\
9& 5153& 2.35& 0.72& 1.59& 1.48& 10581& 7.3& 101& 5.35-16.5& 0.50\\
10& 5139& 2.36& 0.71& 1.74& 1.36& 10953& 5.2& 84.0& 3.45-17& 0.60\\
11& 5129& 2.37& 0.71& 1.87& 1.27& 11223& 3.93& 73.1&2.45-17.5& 0.66\\
12& 5121& 2.37& 0.71& 2.00& 1.19& 11429& 3.07& 65.5& 1.815-18& 0.70\\
\enddata


\tablenotetext{a}{Forward shock expansion parameter.}
\tablenotetext{b}{Free expansion
velocity of ejecta core-envelope boundary.}
\tablenotetext{c}{Time
following explosion when reverse shock enters ejecta core.}
\tablenotetext{d}{Time when blast wave solutions are connected.}
\tablenotetext{e}{Time interval for which ejecta passing through
the reverse shock cools to optically emitting temperatures within
320 years.}
\tablenotetext{f}{Mass of gas that can cool to
optically emitting temperatures within 320 years of explosion.}


\end{deluxetable}

\begin{deluxetable}{cccccccccc}
\tabletypesize{\scriptsize}
\tablecaption{Models $r_b=2.40$ pc, $\rho _{core}=1.22e6$
g cm$^{-3}$s$^3$, and $\rho r_b^2=21$ H atom cm$^{-3}$ pc$^2$\label{tbl-5}}
\tablewidth{0pt}
\tablehead{
\colhead{$n$} & $M_{ej}$ & $E_{51}$ &
\colhead{$v_b\left(320 {\rm yrs}\right)$} &
\colhead{$r_b\left(320 {\rm yrs}\right)$} & \colhead{$\eta $\tablenotemark{a}} &
\colhead{$r_r\left(320 {\rm yrs}\right)$} & \colhead{$r_b/r_r$} &
\colhead{$M_{ej}\left(n-3\right)/n/v_{core}^3$\tablenotemark{b}}\\
 & \colhead{$M_{\sun}$} & \colhead{$10^{51}$ ergs} & \colhead{km s$^{-1}$}
 & \colhead{pc} &  & \colhead{pc} &  & \colhead{g cm$^{-3}$s$^3$}
  }

\startdata
5.5 & 1.72& 5.0 & 5236& 2.391& 0.716& 0.832& 2.87& 1.226e6\\
6& 1.72& 3.85& 5530 & 2.405& 0.752& 0.867& 2.77& 1.219e6\\
7& 1.85& 3.10& 5181& 2.396& 0.707& 1.09& 2.20& 1.226e6\\
8& 1.95& 3.05& 5132& 2.403& 0.699& 1.35& 1.78& 1.225e6\\
9& 2& 3&        5095& 2.398& 0.695& 1.58& 1.52& 1.219e6\\
10& 2.04& 2.99& 5084& 2.401& 0.693& 1.79& 1.34& 1.218e6\\
11& 2.07& 2.98& 5075& 2.402& 0.691& 1.96& 1.23& 1.226e6\\
12& 2.08& 2.97& 5065& 2.402& 0.690& 2.14& 1.12& 1.218e6\\
\enddata


\tablenotetext{a}{Forward shock expansion parameter.}
\tablenotetext{b}{Ejecta core density in velocity space.}

\end{deluxetable}

\begin{deluxetable}{cccccccccc}
\tabletypesize{\scriptsize}
\tablecaption{Models $r_b=2.35$ pc, $\rho _{core}=2.25e6$
g cm$^{-3}$s$^3$, and $\rho r_b^2=14$ H atom cm$^{-3}$ pc$^2$\label{tbl-4}}
\tablewidth{0pt}
\tablehead{
\colhead{$n$} & $M_{ej}$ & $E_{51}$ &
\colhead{$v_b\left(320 {\rm yrs}\right)$} &
\colhead{$r_b\left(320 {\rm yrs}\right)$} & \colhead{$\eta $\tablenotemark{a}} &
\colhead{$r_r\left(320 {\rm yrs}\right)$} & \colhead{$r_b/r_r$} &
\colhead{$M_{ej}\left(n-3\right)/n/v_{core}^3$\tablenotemark{b}}\\
 & \colhead{$M_{\sun}$} & \colhead{$10^{51}$ ergs} & \colhead{km s$^{-1}$}
 & \colhead{pc} &  & \colhead{pc} &  & \colhead{g cm$^{-3}$s$^3$}
  }

\startdata
5.5& 1.815& 4.15& 5134& 2.345& 0.716& 1.28& 1.92& 2.146e6\\
6& 1.815& 2.8& 5395 & 2.347& 0.752& 1.21& 1.94& 2.248e6\\
7& 1.875& 2.15& 5338& 2.349& 0.744& 1.29& 1.82& 2.256e6\\
8& 1.95& 2.05& 5212& 2.349& 0.726& 1.44& 1.63& 2.223e6\\
9& 2& 2& 5153& 2.345& 0.719& 1.59& 1.48& 2.240e6\\
10& 2.035& 1.985& 5130& 2.348& 0.715& 1.73& 1.36& 2.238e6\\
11& 2.065& 1.965& 5105& 2.347& 0.712& 1.85& 1.27& 2.267e6\\
12& 2.075& 1.95& 5085& 2.347& 0.709& 1.96& 1.20& 2.267e6\\
\enddata


\tablenotetext{a}{Forward shock expansion parameter.}
\tablenotetext{b}{Ejecta core density in velocity space.}

\end{deluxetable}

\begin{thebibliography}{}
\bibitem[Aharonian et al.(2001)]{aharonian01} Aharonian, F. et al. 2001, \aap, 370, 112
\bibitem[Akiyama et al.(2003)]{akiyama02}Akiyama, S., Wheeler, J. C., Meier, D. L.,
\& Lichenstadt, I. 2003, \apj, 584, 954
\bibitem[Allende Prieto, Lambert, \& Asplund(2001)]{allende01}Allende Prieto,
C., Lambert, D. L., \& Asplund, M. 2001, \apj, 556, L63
\bibitem[Allende Prieto, Lambert, \& Asplund(2002)]{allende02}Allende Prieto,
C., Lambert, D. L., \& Asplund, M. 2002, \apj, 573, L137
\bibitem[Anders \& Grevesse(1989)]{anders89}Anders, E., \& Grevesse, N. 1989,
Geochim. Cosmochim. Acta, 53, 197
\bibitem[Anderson \& Rudnick(1995)]{anderson95}Anderson, M., \& Rudnick, L. 1995, \apj,
441, 307
\bibitem[Ashworth (1980)]{ashworth80}Ashworth, W. B. 1980, J. Hist. Astron., 11, 1
\bibitem[Berezhko, P\"uhlhofer, \& V\"olk(2003)]{berezhko03} Berezhko, E. G.,
P\"uhlhofer, G., \& V\"olk, H. J. 2003, \aap, 400, 971
\bibitem[Bleeker et al.(2001)]{bleeker01}Bleeker, J. A. M., Willingale, R., van der
Heyden, K., Dennerl, K., Kaastra, J. S., Aschenbach, B., \& Vink,
J. 2001, A\&A, 365, L225
\bibitem[Blondin, Borkowski \& Reynolds (2001)]{blondin01} Blondin, J. M., Borkowski,
K. J., \& Reynolds, S. P. 2001, \apj, 557, 782
\bibitem[Blondin \& Ellison(2001)]{blondin01a}Blondin, J. M., \& Ellison, D. C. 2001,
560, 244
\bibitem[Borkowski et al.(1996)]{borkowski96} Borkowski, K. J., Szymkowiak, A.
E., Blondin, J. M., \& Sarazin, C. L. 1996, \apj, 466, 866
\bibitem[Chevalier \& Fransson(1994)]{chevalier94}Chevalier, R. A., \& Fransson, C. 1994,
\apj, 420, 268
\bibitem[Chevalier(1982)]{chevalier82}Chevalier, R. A. 1982, \apj, 258, 790
\bibitem[Chevalier \& Kirshner(1979)]{chevalier79}Chevalier, R. A., \& Kirshner, R. P.
1979, \apj, 233, 154
\bibitem[Delaney \& Rudnick(2003)]{delaney03} Delaney, T. A., \& Rudnick, L. 2003,
\apj, 589, 818
\bibitem[Fabian et al.(1980)]{fabian80}Fabian, A. C., Willingale, R., Pye, J. P.,
Murray, S. S., \& Fabbiano, G. 1980, \mnras, 193, 175
\bibitem[Favata et al.(1997)]{favata97} Favata, F., Vink, J., Dal Fiume, D.,
Parmar, A. N., Santangelo, A., Mineo, T., Preite-Martinez, A.,
Kaastra, J. S., \& Bleeker, J. A. M. 1997, \aap, 324, L49
\bibitem[Fesen(2001)]{fesen01} Fesen, R. A. 2001, \apjs, 133, 161
\bibitem[Fryer \& Warren(2002)]{fryer02}Fryer, C. L., \& Warren, M. S. 2002, \apj,
574, L65
\bibitem[Fryer \& Warren(2003)]{fryer03}Fryer, C. L., \& Warren, M. S. 2003, in
preparation
\bibitem[Fryer \& Heger(2000)]{fryer00}Fryer, C. L., \& Heger, A. 2000, \apj, 541, 1033
\bibitem[Garcia-Segura, Langer, \& Mac Low(1996)]{garcia96}Garcia-Segura, G.,
Langer, N., \& Mac Low, M.-M. 1996, \aap, 316, 133
\bibitem[Gotthelf et al.(2001)]{gotthelf01}Gotthelf, E. V., Koralesky, B., Rudnick, L.,
Jones, T. W., Hwang, U., \& Petre, R. 2001, \apj 552, L39
\bibitem[Grevesse \& Sauval(1998)]{grevesse98}Grevess, N., \& Sauval, A. J. 1998,
Space Science Reviews, 85, 161
\bibitem[Holweger(2001)]{holweger01}Holweger, H. 2001 in AIP Conference Proceedings 598,
Solar and Galactic Composition, ed. R. F. Wimmer-Schweingruber, p23
\bibitem[Hwang \& Laming(2003)]{hwang03}Hwang, U., \& Laming, J. M. 2003, \apj, submitted
\bibitem[Iyudin et al.(1994)]{iyudin94}Iyudin, A. F., et al. 1994, \aap, 284, L1
\bibitem[Iyudin et al.(1997)]{iyudin97}
Iyudin, A. F., Diehl, R., Lichti, G. G., et al. 1997, ESA SP-382,
37
\bibitem[Kamper \& van den Bergh(1976)]{kamper76}Kamper, K., \& van den Bergh, S. 1976,
\apjs, 32, 351
\bibitem[Khokhlov et al.(1999)]{khokhlov99}Khokhlov, A. M., H\"oflich, P. A.,
Oran, E. S., Wheeler, J. C., Wang, L., \& Chtchelkanova, A. Yu. 1999, \apj, 524, L107
\bibitem[Klein, McKee \& Colella(1994)]{klein94} Klein, R. I., McKee, C. F., \& Colella,
P. 1994, \apj, 420, 213
\bibitem[Klein et al.(2003)]{klein03}Klein, R. I., Budil, K. S., Perry, T. S., \&
Bach, D. R. 2003, \apj, 583, 245
\bibitem[Koralesky et al.(1998)]{koralesky98}Koralesky, B., Rudnick, L., Gotthelf, E. V.,
\& Keohane, J. W. 1998, \apj, 505, L27
\bibitem[Kassim et al.(1995)]{kassim95}Kassim, N. E., Perley, R. A., Dwarakanath, K. S.,
\& Erickson, W. C. 1995, \apj, 455, L59
\bibitem[Lamers \& Nugis(2002)]{lamers02}Lamers, H. J. G. L. M., \& Nugis, T. 2002,
\aap, 395, L1
\bibitem[Laming(2001a)]{laming01a} Laming, J. M. 2001a, \apj, 546, 1149
\bibitem[Laming(2001b)]{laming01b}Laming, J. M. 2001b, \apj, 563, 828
\bibitem[Laming \& Grun(2002)]{laming02}Laming, J. M., \& Grun, J. 2002, \prl, 89, 125002
\bibitem[Laming \& Grun(2003)]{laming03}Laming, J. M., \& Grun, J. 2003, Physics of
Plasmas, in press
\bibitem[Massey, DeGioia-Eastwood, \& Waterhouse(2001)]{massey01}Massey, P.,
DeGioia-Eastwood, K., \& Waterhouse, E. 2001, \aj, 121, 1050
\bibitem[Mazzotta et al.(1998)]{mazzotta98}Mazzotta, P., Mazzitelli, G., Colafranceso, S.,
\& Vittorio, N. 1998, \aaps, 133, 403
\bibitem[McKee \& Cowie(1975)]{mckee75}McKee, C. F., \& Cowie, L. L. 1975, \apj, 195, 715
\bibitem[McKee \& Truelove(1995)]{mckee95}McKee, C. F., \& Truelove, J. K. 1995,
Phys. Rep., 256, 157
\bibitem[Nagataki et al.(1998)]{nagataki98}Nagataki, S., Hashimoto, M., Sato, K.,
Yamada, S., \& Mochizuki, Y. 1998, \apj, 492, L45
\bibitem[Poludnenko, Frank, \& Blackman(2001)]{poludnenko01}Poludnenko, A. Y., Frank,
A., \& Blackman, E. G. 2001, \apj, 576, 832
\bibitem[Reber(1944)]{reber44} Reber, G. 1944, \apj, 100, 279
\bibitem[Reed et al.(1995)]{reed95} Reed, J. E., Hester, J. J., Fabian, A. C.,
\& Winkler, P. F. 1995, \apj, 440, 706
\bibitem[Reichart \& Stephens(2000)]{reichart00}Reichart, D. E., \& Stephens, A. W.
2000, \apj, 537, 904
\bibitem[Rothschild \& Lingenfelter(2003)]{rothschild03}Rothschild, R. E., \&
Lingenfelter, R. E. 2003, \apj, 582, 257
\bibitem[Ryle \& Smith(1948)]{ryle48} Ryle, M., \& Smith, F. G. 1948, Nature, 162, 462
\bibitem[Sgro(1975)]{sgro75}Sgro, A. 1975, \apj, 197, 621
\bibitem[Spitzer(1978)]{spitzer78} Spitzer, L. Jr., Physical Processes in the
Interstellar Medium, (New York: Wiley)
\bibitem[Summers \& McWhirter(1979)]{summers79}Summers, H. P., \& McWhirter, R. W. P.,
J. Phys. B., 12, 2387
\bibitem[Sutherland \& Dopita(1995)]{sutherland95}Sutherland, R. S., \& Dopita, M. A.
1995, \apj, 439, 381
\bibitem[Thorstensen, Fesen, \& van den Bergh(2001)]{thorstensen01}Thorstensen, J. R.,
Fesen, R. A., \& van den Bergh, S. 2001, \aj, 122, 297
\bibitem[Timmes et al.(1996)]{timmes96} Timmes, F. X., Woosley, S. E., Hartmann, D.
H., \& Hoffman, R. D. 1996, ApJ, 464, 332
\bibitem[Truelove \& McKee(1999)]{truelove99}Truelove, J. K., \& McKee, C. F. 1999, \apjs,
120, 299
\bibitem[van den Bergh \& Kamper(1985)]{vandenbergh85}van den Bergh, S., \& Kamper, K.
1985, \apj, 293, 537
\bibitem[van den Bergh(1971)]{vandenbergh71}van den Bergh, S. 1971, \apj, 165, 457
\bibitem[Vink \& Laming(2003)]{vink02}Vink, J., \& Laming, J. M. 2003, \apj, 584, 758
\bibitem[Vink et al.(2001)]{vink01}Vink, J., Laming, J. M., Kaastra, J. S.,
Bleeker, J. A. M., Bloemen, H., \& Oberlack, U. 2001, \apj, 560,
L79
\bibitem[Vink et al.(1998)]{vink98} Vink, J., Bloemen, H., Kaastra, J. S., \&
Bleeker, J. A. M. 1998, \aap, 339, 201
\bibitem[Vink, Kaastra, \& Bleeker(1996)]{vink96} Vink, J., Kaastra, J. S., \&
Bleeker, J. A. M. 1996, \aap, 307, L41
\bibitem[Wang \& Chevalier(2001)]{wang01}Wang, C.-Y., \& Chevalier, R. A. 2001, \apj,
549, 1119
\bibitem[Wang \& Chevalier(2002)]{wang02}Wang, C.-Y., \& Chevalier, R. A. 2002, \apj,
574, 155
\bibitem[Willingale et al.(2003)]{willingale02}Willingale, R., Bleeker, J. A. M., van
der Heyden, K. J \& Kaastra, J. S. 2003, \aap, 398, 1021
\bibitem[Woosley, Langer, \& Weaver(1993)]{woosley93}Woosley, S. E., Langer, N., \&
Weaver, T. A. 1993, \apj, 411, 823

\end{thebibliography}
\end{document}